\documentclass[11pt]{article}

\usepackage{lipsum} 

\usepackage{bm}
\usepackage{amsmath,amsfonts,amsthm}
\usepackage{esdiff}  
\usepackage{booktabs}  
\usepackage{url}  
\usepackage{cleveref}  
	\crefname{equation}{equation}{equations}
	\crefname{figure}{figure}{figures}	
	\crefname{table}{table}{tables}
\usepackage[skip=1.5pt,font=small]{caption}

\usepackage{caption}
\usepackage{bbm}

\usepackage[usenames,dvipsnames,svgnames,table]{xcolor}

\usepackage{graphicx}
\usepackage{hyperref}  

\usepackage[sc]{mathpazo} 
\usepackage[T1]{fontenc} 
\usepackage{microtype} 

\usepackage{multicol} 
\usepackage[margin={1cm,1cm}]{geometry}
\usepackage{booktabs} 
\usepackage{float} 
\usepackage{hyperref} 

\usepackage{lettrine} 
\usepackage{paralist} 

\usepackage{abstract} 

\usepackage{titlesec} 
\renewcommand\thesection{\Roman{section}} 
\renewcommand\thesubsection{\Alph{subsection}} 
\titleformat{\section}[block]{\large\scshape\centering\bfseries}{\thesection.}{1em}{} 

\titleformat{\subsection}[block]{\scshape\centering}{\thesubsection.}{1em}{} 

\usepackage{fancyhdr} 
\DeclareCaptionFormat{myformat}{#1#2#3\hrulefill}
\usepackage{authblk}
\makeatletter
\renewcommand\AB@affilsepx{, \protect\Affilfont}
\makeatother
\title{\vspace{-15mm}\fontsize{16pt}{16pt}\selectfont\textbf{Pipe-cleaner Model of Neuronal Network Dynamics}} %
\author[1]{Eve Armstrong\thanks{earmstrong@ucsd.edu}}
\affil[1]{BioCircuits Institute, University of California, San Diego, La Jolla, CA 92093-0374} 
\date{(Dated: April 1, 2016)\vspace{-2ex}}
\renewcommand\Affilfont{\itshape\small}

\begin{document}

\maketitle 



\begin{abstract}
We present a functional model of neuronal network connectivity in which the single architectural element is the object commonly known in handicraft circles as a pipe cleaner.  We argue that the dual nature of a neuronal circuit - that it be at times highly robust to external manipulation and yet sufficiently flexible to allow for learning and adaptation - is embodied in the pipe cleaner, and thus that a pipe cleaner framework serves as an instructive scaffold in which to examine network dynamics.  Regarding the dynamics themselves: as pipe cleaners possess no intrinsic dynamics, in our model we attribute the emergent circuit dynamics to magic.  Magic is a strategy that has been largely neglected in the neuroscience community, and may serve as an  illuminating comparison to the common physics-based approaches.  This model makes predictions that it would be really awesome to test experimentally.  Moreover, the relative simplicity of the pipe cleaner - setting aside the fact that it comes in an overwhelming variety of colors - renders it an excellent theoretical building block with which to create simple network models.  Also, they are incredibly cheap when bought wholesale on Amazon.
\end{abstract}

\section{\\INTRODUCTION}
\begin{multicols}{2}
The use of pipe cleaners\footnote{A pipe cleaner is a piece of wire lined with tufted fiber.  It is available for private and wholesale purchase throughout the United States and most of the world.  Originally invented for cleaning pipes, the pipe cleaner's uses have broadened in recent times to include the cleaning of small spaces of unspecified nature, and more commonly in creative endeavors (Wikipedia, "Pipe Cleaner", 2016).} in scientific investigations has a rich history in elementary school classrooms and summer camps throughout the world.  The pipe cleaner's complementary traits of pliability and sturdiness render it more versatile than other common construction materials such as popsicle sticks and drinking straws.  Its availability in a variety of colors\footnote{Some even glow in the dark (Homecrafts 2016).} has further enhanced its popularity among the scientifically inclined under-age-ten population (Ansberry 2010).  Use of the pipe cleaner in more advanced scientific arenas, however, has been explored minimally.  We aim to remedy this oversight, specifically in the field of neuroscience, where we shall argue that axons and synapses possess distinct pipe-cleaner-esque morphologies.  For this reason, we expect that more active exploration of pipe cleaner modeling would be pivotal to the field and would serve as an instructive comparison to the more common computational modeling techniques that are grounded in "reality".

Previous models have considered neuronal networks in terms of complex nonlinear dynamical systems (e.g. Izhikevich 2007).  These models have succeeded in capturing some qualities of the observed biological systems; however, it is clear that no such model to date is perfect to infinite precision.  Despite this limitation, which is inherent in discretized physical models, physics is still considered to be one of the few acceptable methods of predicting the future behavior of a biological network.\footnote{The other commonly-accepted method is Classifying Things Without Knowing Why.}  This convention persists despite a broad body of evidence for both stability and versatility in pipe cleaners (Wikipedia, "Pipe cleaners", 2016) - signs that they possess promise as a potential alternative modeling technique. 

We depart from earlier modeling efforts by considering a neural circuit, and the individual neurons therein, to be a collection of distinct pipe cleaners of identical composition.  In this way, we aim to capture the two salient features of a neuronal network: first that it generate stable activity that is relatively robust to external manipulation, and secondly that it possess the flexibility necessary for learning and adaptation.  We note the same potentially bi-modal traits in the pipe cleaner, which can be wrapped in a manner so tight that the resulting attachment becomes extremely difficult to re-adjust, and which can also bend in any direction, at any location along the length, with a bend radius of essentially zero.  Finally, we suggest that the model network dynamics themselves can be achieved via magic, an approach that might serve as an instructive comparison to physics-based approaches.  Indeed, it should be noted that real neurons \textit{in vitro} often display spontaneous spiking behavior (e.g. Daou et al. 2013; Bennett \& Wilson 1999), indicating that they can at times behave as perpetual motion machines, thereby unequivocally violating the laws of physics.
\end{multicols}

\section{\\METHODS}
\begin{multicols}{2}
\subsection{Materials}
We obtained 108 pipe cleaners termed "chenille stems" by $Darice^{\circledR}$ (Darice, Inc. 2016) that were designated as hobby-and-craft items for use by investigators over the age of seven.  These objects were made of two lengths of wire twisted together and trapping short lengths of polyester fibre between them.  During production, the pipe cleaners had been wound onto spools, where they were subsequently cut to lengths of 30.5 cm, with a diameter of 1 mm.\footnote{This corresponds to an American Wire Gauge of roughly 18.  We were unable to ascertain from the package label the type of metal used; we hypothesize that it is steel, as copper and aluminum are too soft at that gauge and would be prone to breaking.}  The polyester fibre of each pipe cleaner was dyed in two-colored stripes.  The top panel of Figure 1 shows all pipe cleaners used in this study.

\subsection{Neurons}
We based our constituent neuron models on two fundamental experimental observations: 1) axons and dendrites look like pipe cleaners; 2) soma bodies look a little bit like balled-up pipe cleaners (You 2016).  Armed with this evidence, we constructed our model neurons as follows.

We created a total of six five-compartment neuron models: three excitatory and three inhibitory.  Each neuron consisted of a soma body, one axon, and three dendritic projections.  To construct the soma, we bunched up one pipe cleaner into a ball.  Each axon consisted of one full-length pipe cleaner with small pieces of other pipe cleaners extending from it.  Each dendrite consisted of a half-length of pipe cleaner with multiple branches.  To distinguish compartments, we used green and purple pipe cleaners for the axons and dendrites, respectively; the soma bodies were red for excitatory and yellow for inhibitory cells.  The middle panel of Figure 1 shows these constituents; the bottom panel shows one representative five-compartment excitatory neuron.
\subsection{Connectivity}
To choose our connectivity design, we sought to emulate a technique in which connectivity is drawn randomly from a probability distribution, where the distribution is set by various learning rules (e.g. Huerta \& Rabinovich 2004; Rajan \& Abbott 2006).  To simulate this strategy,
\begin{figure}[H]
\centering
  \includegraphics[width=80mm]{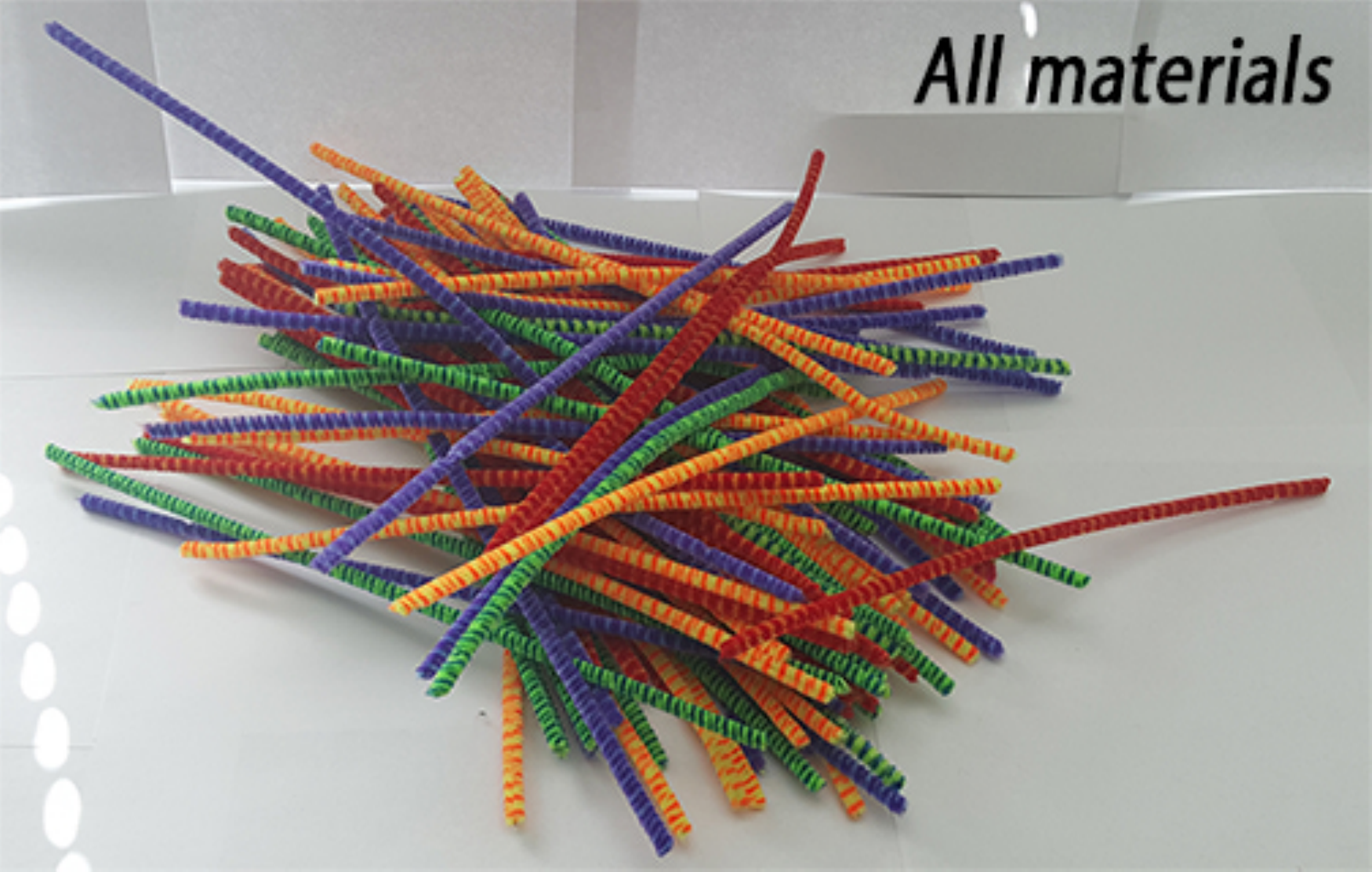}\\
  \vspace{3mm}
  \includegraphics[width=80mm]{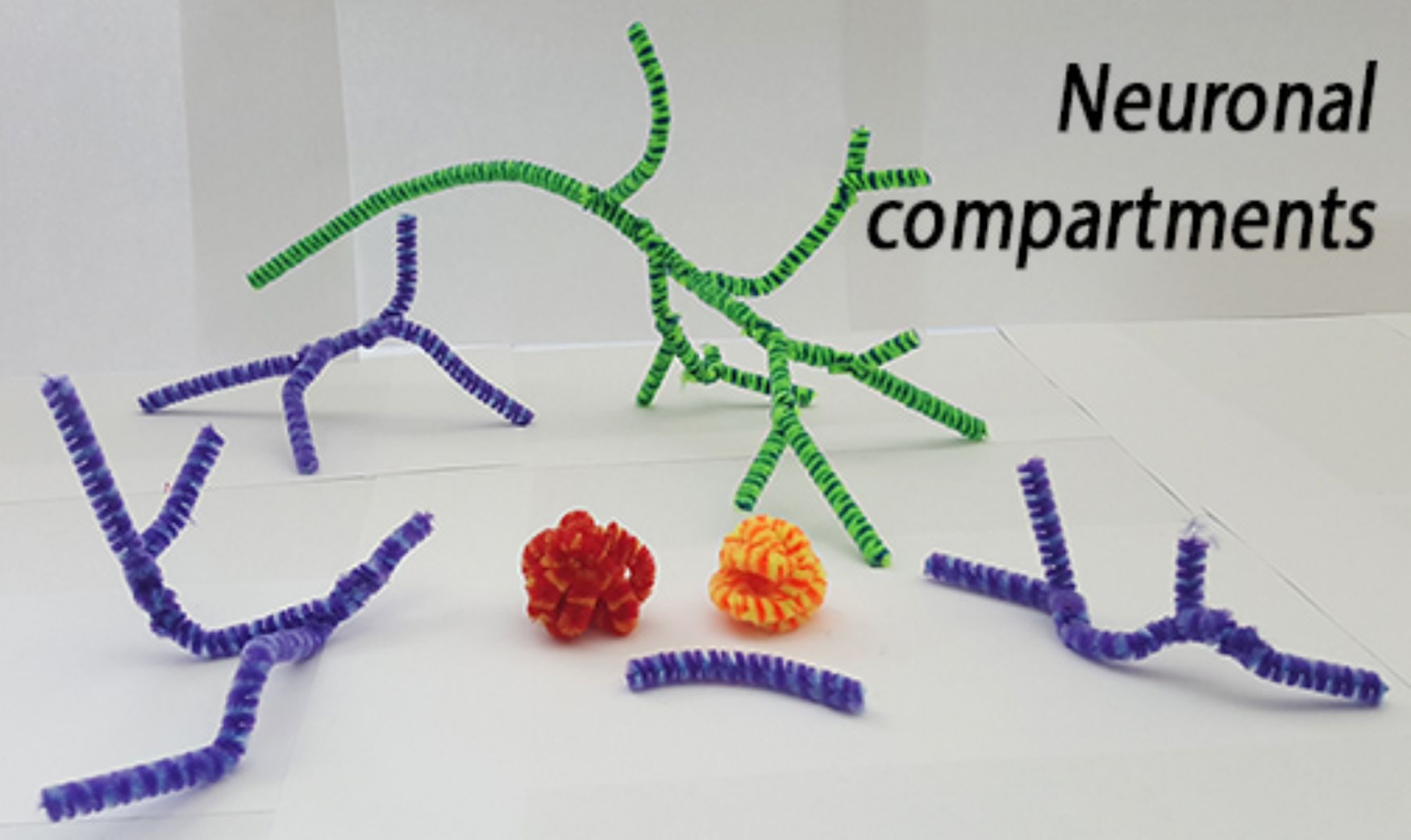}\\
  \vspace{3mm}
  \includegraphics[width=80mm]{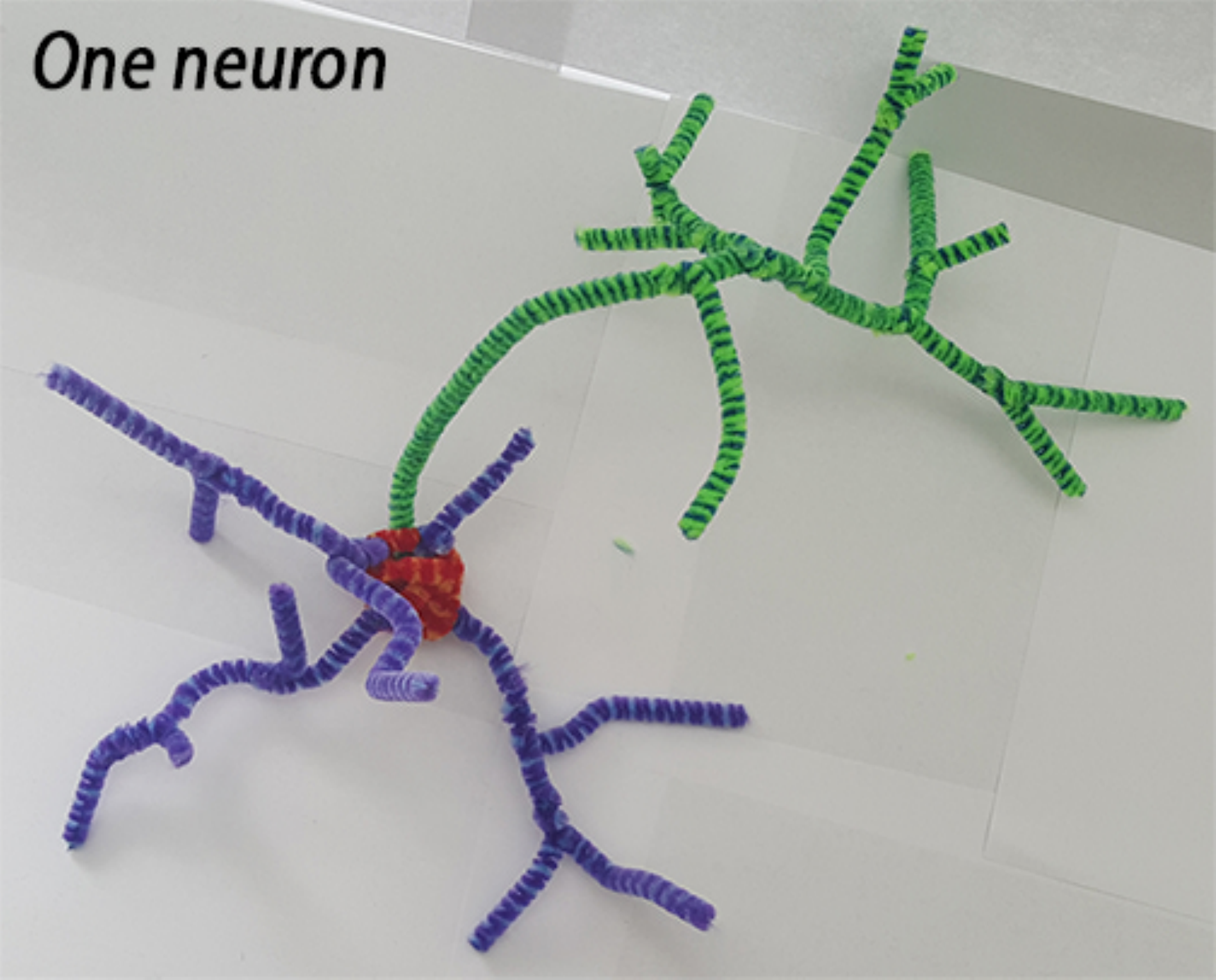}
  \caption{\textbf{Pipe cleaners used in model construction.}  \textit{Top}: All of the pipe cleaners, identical in size and morphology.  \textit{Middle}: The component axons (green), dendritic trees (purple), excitatory soma bodies (red), and inhibitory soma bodies (yellow), respectively.  \textit{Bottom}: One complete neuron model.}
\end{figure}
\noindent
we stuffed the six model pipe cleaner neurons into a plastic bag and vigorously shook the bag.  Whichever neurons were subsequently found to be hopelessly intertwined upon removal from the bag were taken to be synaptically connected.  This procedure was carried out subject to the following \textbf{learning rules}: 
\begin{itemize}
  \item i) all neurons must be stuffed into the bag; 
  \item ii) the intentional connecting of any neurons prior to bag closing is not permitted;
  \item iii) the bag must be double-bagged to prevent tearing. 
\end{itemize}  

Figure 2 shows the outer bag used to effect the connectivity; the inner bag, for all intents and purposes, is identical.  These bags were obtained from Ralph's Grocery Story at 8657 Villa La Jolla Dr, La Jolla, CA 92037.  They were manufactured by $QuikMate^{\circledR}$  (Novelex Corporate 2016) and allegedly contain a minimum of 25 per cent recycled material of an unspecified nature. 

\subsection{Invoking dynamics}
We investigated two alternative approaches to invoke the pipe cleaner network dynamics: 1) equations of motion coupled with magic, and 2) just magic. \\

\noindent
\textbf{\textit{Using first-order equations of motion and magic}}\\

We first wrote down the equations of motion and showed them to the circuit that emerged from the shaken plastic bag.  The implicit assumption behind this approach was that the power of suggestion might motivate the pipe cleaners to motion.  

For each neuron, we considered the dynamics to be uniform across all five neuronal compartments, and we assigned a Hodgkin-Huxley-type model with sodium, potassium, and leak currents only.  The time evolution of the membrane potential V(t) of each neuron is expressed as:
\begin{align*} 
  \diff{V_{i}(t)}{t} &= \frac{1}{C}(I_{L,i}(t) + I_{Na,i}(t) +  I_{K,i}(t) + \sum_{j \neq i}I_{syn,ij}(t)
\end{align*}
\begin{align}
        &+ I_{background})
\end{align}
\noindent
where $I_{background}$ represents ambient background excitation, and the ion channel currents are: 
\begin{align*} 
  I_{L,i}(t) &= g_{L}(E_{L} - V_i(t))\\ 
  I_{Na,i}(t) &= g_{Na,i} m(t)^3 h(t) (E_{Na} - V_i(t)) \\
  I_{K,i}(t) &= g_{K,i} n(t)^4 (E_K - V_i(t)).
\end{align*}
\begin{figure}[H]
\centering
  \includegraphics[width=80mm]{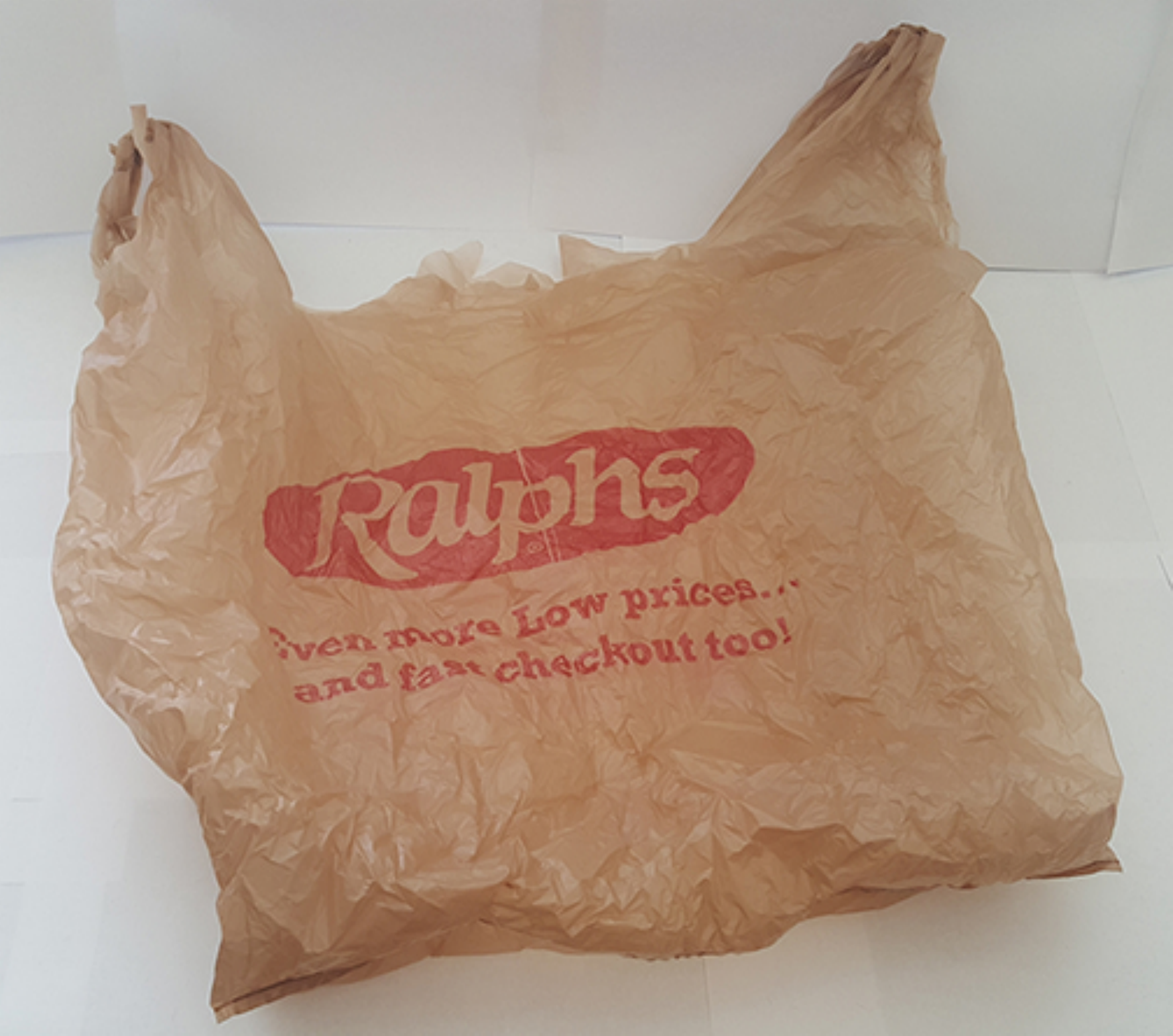}\\
  \caption{\textbf{The outer of two double-bagged plastic bags used to effect the random neuronal connectivity.}   The pipe cleaner neurons were stuffed into this bag whereafter the bag was shaken vigorously for 27 seconds.  Whichever neurons were observed to be hopelessly intertwined upon removal from the bag were then considered to be synaptically connected, subject to the learning rules specified in the text.  The bag was obtained from Ralph's Grocery Story at 8657 Villa La Jolla Dr, La Jolla, CA 92037.  It was fabricated by $QuikMate^{\circledR}$ (Novolex Corporate 2016), and it allegedly contains a minimum of 25 per cent recycled material of an unspecified nature.}
\end{figure}
\noindent
The gating variables m, h, and n satisfy:
\begin{align*} 
  \diff{a_i}{t}(t) &= (a_{\infty}(V_i(t)) - a_i(t))/\tau_a(V_i(t)); \\
  a_{\infty}(V_i) &= 0.5 [1 + \tanh((V_i - \theta_{a,i})/\sigma_{a,i})]\\
  \tau_a(V_i) &= t_{a0} + t_{a1}[1 - \tanh^2((V_i - \theta_{a,i})/\sigma_{a,i})];\\
  a(t) &= m(t),h(t),n(t). 
\end{align*} 
\noindent
We define a particular neuron as "excitatory" or "inhibitory" via resting potential $E_L$ and synaptic reversal potential $E_{syn,ij}$.  Definitions for all other terms can be found in plenty of other places (e.g. Gibb, Gentner, \& Abarbanel 2009).  We do not list the parameter values because we doubt that the reader wants to see them.\footnote{Unless, of course, the reader is an experimental neurophysiologist, in which case we would be happy to email you the table.}

The synaptic current is expressed as: 
\begin{align*} 
  I_{syn,ij} &= g_{ij} s_{ij}(t)(E_{syn,ij} - V_{i}(t))\\  
\end{align*}
\noindent
where we have taken the formalism of Destexhe \& Sejnowski (2001) and Destexhe et al. (1994); see those sources for the form of the synapse gating variable $s_{ij}$. \\
\newpage
\noindent
\textbf{\textit{Using just magic}}\\

As an alternative approach, we dispensed entirely with the equations of motion and retained only the magical component of our dynamics-eliciting strategy.  To adhere as closely as possible to an electrically-based Hodgkin-Huxley model, we first explored Fulgurkinesis, defined as mastery over lightning  (Telekinetic Wikia 2016).  We also experimented with Telekinesis, the ability to move objects with one's mind (Wikipedia, "Telekinesis", 2016). 
\end{multicols}

\section{\\RESULTS}
\begin{multicols}{2}
\subsection{Emergent connectivity after shaking up neurons in a plastic bag}
Figure 3 shows the six neurons prior to bag-shaking (top) and the emergent circuit following bag-shaking (bottom).  The result is a highly interconnected web, whose connectivity matrix is listed in Table 1.  

One observation is particularly noteworthy: the soma body of one of the excitatory neurons fell off (this neuron is denoted "Cell 2" in Table 1).  We still chose to record its synaptic connections in the matrix, because while its axons and dendrites separated from the soma, they became tightly interwoven with nearly all of the other neurons - and with each other (with nine self-connections).  In addition, we considered this result to be a unique opportunity to observe possible effects of a soma-less neuron on network dynamics, as such an observation has not been noted in the literature.  
\setlength{\tabcolsep}{1.5pt}
\begin{table}[H]
\centering
\begin{tabular}{ l c c c c c c } \toprule
 \textit{Cell} & \textit{1} & \textit{2} & \textit{3} & \textit{4} & \textit{5} & \textit{6} \\\midrule 
 \textit{1} & 2  & 2  & 4  & 1  & 3  & 3 \\
 \textit{2} & 3  & 9  & 3  & 2  & 2  & 2  \\
 \textit{3} & 3  & 2  & 2  & 4  & 1  & 3 \\
 \textit{4} & 0  & 2  & 2  & 2  & 3  & 3\\
 \textit{5} & 3  & 4  & 1  & 2  & 4  & 3 \\
 \textit{6} & 1  & 0  & 2  & 3  & 3  & 2 \\\bottomrule
\end{tabular}\\
\caption{\textbf{Synapse strengths $g_{ij}$ for the six neurons after the vigorous bag shaking}.  The numbers are dimensionless, where 1, 2, and 3 indicate one, two, or three connections between the pair of neurons, respectively.  \textit{Notation}: The value of 2 in [row 1, column 2] corresponds to $g_{12}$: the connection entering Cell 1 from Cell 2.}
\end{table}
\subsection{Eliciting network dynamics}
\noindent
\textbf{\textit{Failed attempts using the equations of motion}}\\

We first tried to motivate the dynamics of our pipe cleaner model via the equations of motion.  We wrote the equations in legible handwriting on large pieces of paper and taped the papers onto a wall to which all neurons had a clear line-of-sight.  Next, to simulate the external
\begin{figure}[H]
\centering
  \includegraphics[width=90mm]{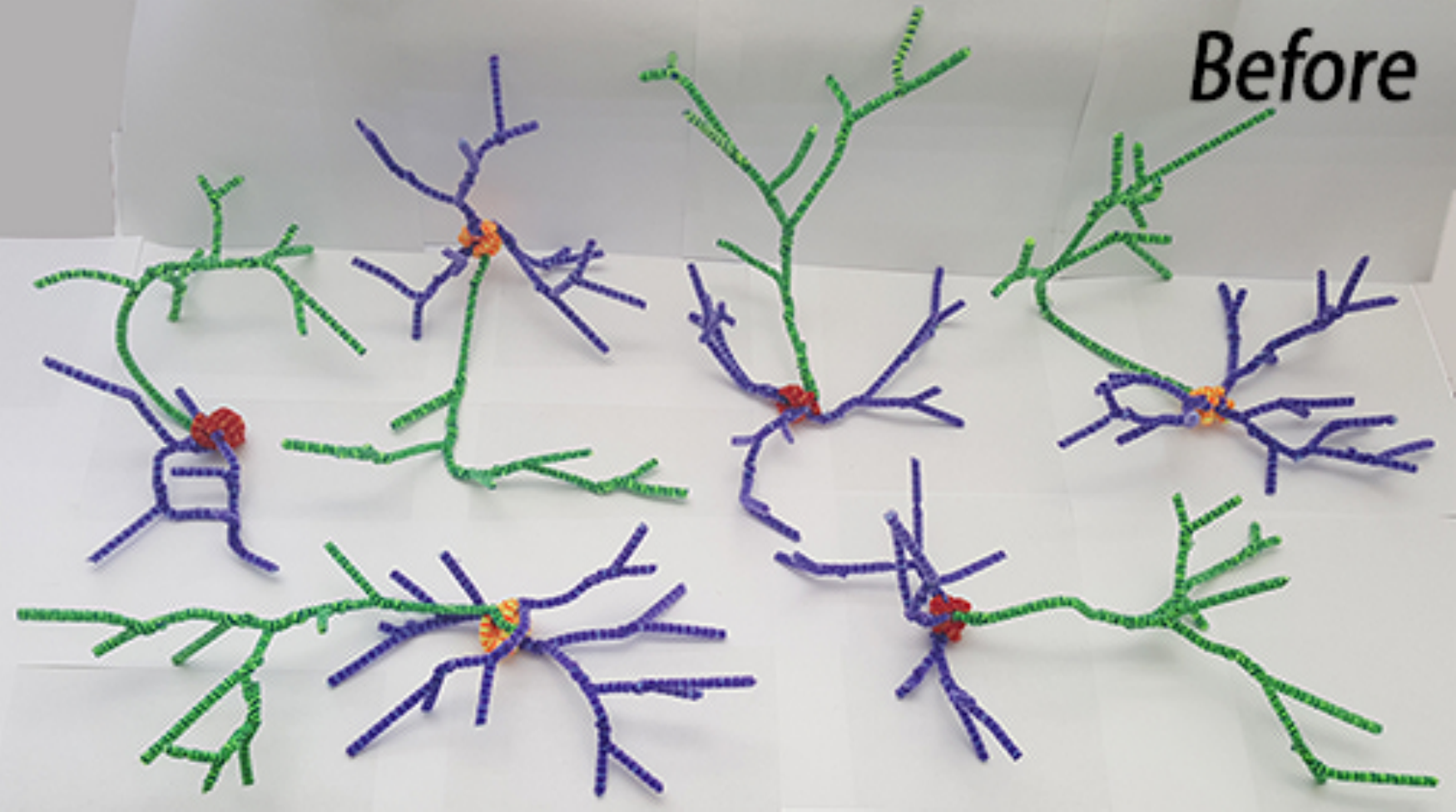}\\
  \vspace{3mm}
  \includegraphics[width=90mm]{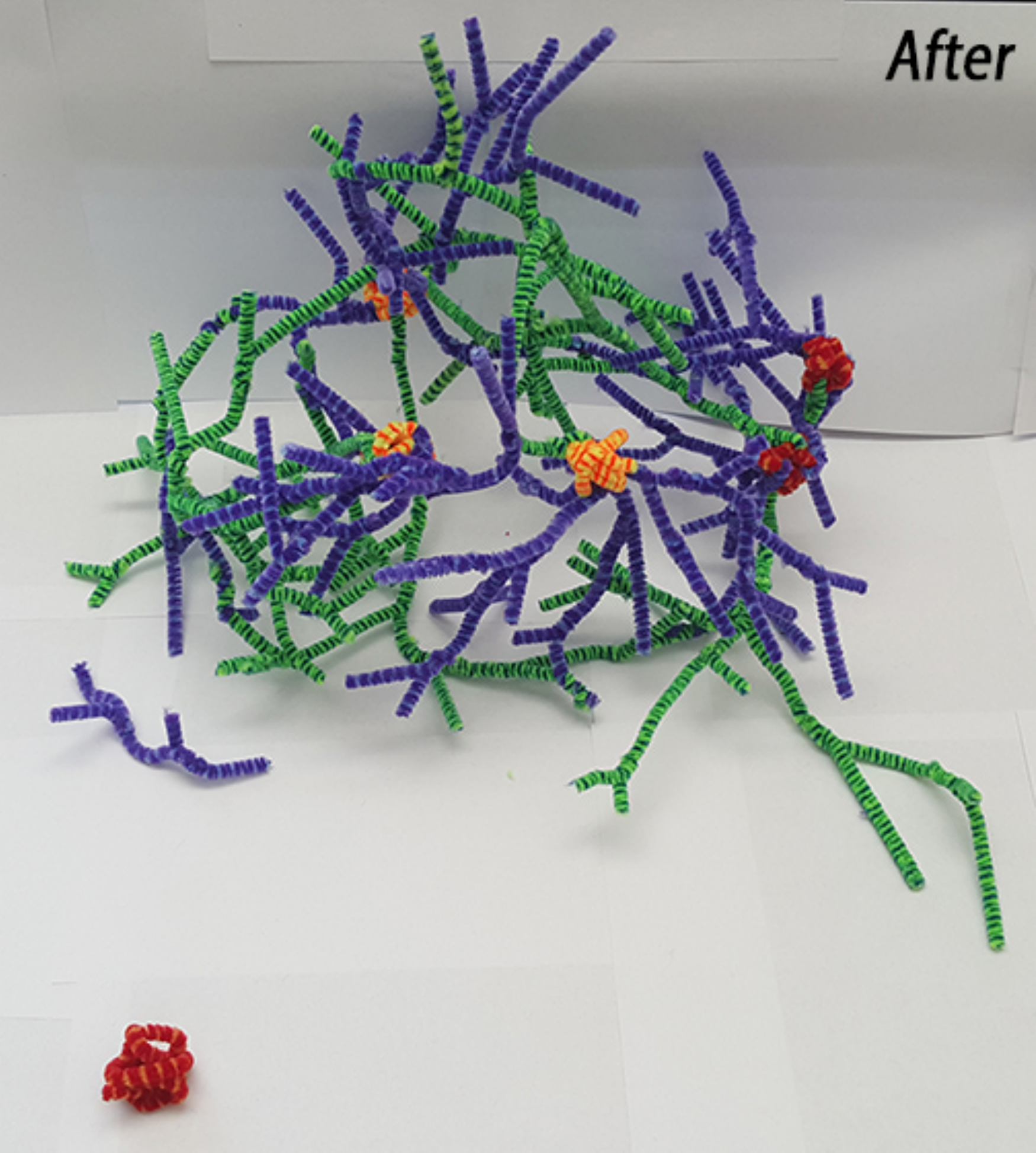}
  \caption{\textbf{Effecting random connectivity by stuffing all neurons into a bag and shaking the bag.}  \textit{Top}: The six individual neurons prior to being stuffed into the bag.  \textit{Bottom}: The resultant connectivity among the neurons following a vigorous 27-second bag-shaking.  Table 1 shows the connectivity matrix.  Note that the soma body of one of the excitatory neurons (Cell 2 in Table 1) has fallen off.}
\end{figure}
\noindent
$I_{background}$ term of Equation 1, we dangled six 60-Watt light bulbs from the ceiling above the circuit, each positioned directly above one of the six soma bodies to within eight centimeters.  Then we waited to see whether the pipe cleaners would take the hint and spontaneously spring to life.  

Our motivations for this approach were threefold.  First, as noted earlier, some neurons in the central nervous system exhibit spontaneous spiking activity \textit{in vitro} in the absence of external stimuli.  If biological neurons are able to act as perpetual motion machines, thereby violating the laws of physics, why not their pipe cleaner counterparts?  Secondly, many inanimate objects in movies, particularly horror films, have come to life.  For example: in "The Lift", the animated object is an elevator (\textit{The Lift}, 1983); in \textit{The Refrigerator} (1991): a refrigerator; in \textit{Death bed ... The Bed that Eats!} (1977): a bed; in \textit{Maximum Overdrive} (1986): a vending machine.  And, as is commonly known, art imitates real life.  Finally, note Gumby.  Gumby was a freakish humanoid made of green clay.  How in the world did that work?  We have yet to find any authoritative reference to answer this question.

We waited for the pipe cleaners to spring to action, several times experimenting with the light bulb fixtures.  We were unable to evoke spiking activity in any of the pipe cleaners, regardless of the light bulb configuration.\\

\noindent
\textbf{\textit{Mild success using magic}}\\

Next we eliminated the equations of motion from consideration and sought to determine whether we could elicit dynamical behavior by directing our efforts entirely to magic.  To this end, we first briefly explored Fulgurkinesis: mastery over lightning.  The aim here was to retain the electrically-based feature captured in the Hodgkin-Huxley model.  The only strategy we hit upon to effect mastery over lightning was simply to think hard about lightning and hope that it would strike our circuit.  This strategy did not work.  We note, however, the possibly-mitigating circumstance that our lab resides in San Diego, California, and that lightning has not struck here at any time in recent memory.

Finally we applied telekinesis: the ability to move objects with one's mind.  We interpreted this strategy as wishing really really hard that the neurons would move.  We embellished this basic technique by staring menacingly at the neurons, yelling at the neurons, and uncontrollably sobbing.  After seven minutes, one of the dendritic projections of Neuron 5 fell off the table onto the floor.  Encouraged by this progress, we enthusiastically continued our efforts for 15 more minutes before getting tired of it and going out bowling.

To summarize: given that one of the pipe cleaners was endowed with motion for a brief (2.1-second) period, there appears to be promise in our Magic-Only approach to effecting network dynamics.  This tentative finding merits a detailed follow-up study.

\subsection{Model scalability}
  
Our pipe cleaner model is generalizable to consist of an arbitrary number of pipe cleaner neurons.  The rules are as follows: 
\begin{itemize}
  \item The pipe cleaners must be sufficiently fuzzy, to facilitate adhesion for synaptic connections; we recommend a fuzz radius of no fewer than 6 mm;
  \item When incorporating more than \textasciitilde ten pipe cleaner neurons into the circuit, use a larger bag;
  \item Do not trick your color-blind collaborator by switching the labeling of his red and green pipe cleaners on the very morning of his APS presentation, to embarrass him in front of hundreds of people.  We won't name names; you know who you are.  
\end{itemize}  
\end{multicols}  

\section{\\DISCUSSION}
\begin{multicols}{2}
\subsection{The promise of magical network dynamics}
The mild success of magic as an underlier of our model network dynamics gives us pause.  We have been unable to find in the recent (post 1800's) literature a record of magic being used in either neuroscience or in the modeling of complex dynamical systems in general.  Given the indications that magic effected the falling of Neuron 5's dendrite off the table onto the floor, it is time to remedy that oversight.

Admittedly, the falling of Neuron 5's dendrite was a transient event.  We were unable to use the magical approach to elicit continuous and long-term dynamical behavior.  There exists some historical precedent, however, suggesting that we might not have persevered in our efforts for a sufficient duration.  Namely: the importance of patience in effecting magical transformations has been documented.  

In the early 16th century, Jan Baptiste van Helmont, a specialist in spontaneous generation, cited a 21-day waiting period for the effective transformation of wheat and dirty clothing into mice (Pasteur 1864).  The German weaver Rumpelstilzchen also deferred to a requisite timescale for effecting such transformations.  By the account of Grimm \& Grimm (1812), a successful effort to spin straw into gold demanded of Rumpelstilzchen a full night of uninterrupted concentration. 

Magic possesses yet another useful quality: it avoids the cumbersome requirement to use mathematical equations that, when discretized and coded, do not always yield the dynamics that one would prefer.  Rather, often one's code yields results that do not support the argument that one had already decided to make\footnote{Indeed, it can seem as though one has created a Frankenstein Monster with a mind of its own - which, one might recall, was made of reanimated dead tissue.}, a situation that can be quite vexing.  One must decide whether to change one's conclusion for consistency with the result or instead to ignore the result and continue to edit the code until its output becomes consistent with one's preconceived conclusion.  The optimal choice is not always obvious.

Magic has the potential to render the problem vastly simpler.  Indeed, we envision that in the best-case scenario, one might simply decide that pipe cleaners possess whatever dynamics one desires.  In this framework, Science is whatever we would like it to be.

\subsection{Effects of a soma-less neuron on circuit dynamics}
What was the impact of the partially disembodied neuron that emerged in our randomly-connected circuit?  As the circuit did not do much of anything, we were unable to examine possible modulating factors in any detail.  Thus, any suggestions of causal relationships between soma-less-ness and network dynamics would be speculative.  This emergent network property merits a more detailed follow-up.

\subsection{Model predictions and suggested experiments}
Here we describe possible tests of our model in a laboratory setting.

\begin{itemize}
  \item \textit{To test whether the equations of motion really are sufficient to effect the pipe cleaner dynamics:} \textbf{Set the lab on fire.}  It is possible that the dangling 60-Watt light bulbs did not provide sufficient background excitation to serve as an adequate $I_{background}$ term in Equation 1.  
  \item \textit{To test whether the one soma-less neuron was responsible for the failure of magic to elicit long-term continuous dynamics:} \textbf{Remove the soma bodies from \textit{all} of the neurons, and repeat the full set of experiments.}  It is possible that the soma-less neuron, never before witnessed in real biological circuits, was alone responsible for the dearth of network dynamics.  If soma-less-ness is responsible for the lack of dynamics, then by removing the somas from \textit{all} neurons, one should find that \textit{none} of the dendrites fall off the table onto the floor.  
  \item \textit{To test whether the neurons were just not in the mood to fire:} \textbf{Bring in a motivational speaker such as Oprah Winfrey to coach the neurons into feeling good about themselves.}  It is possible that our incessant 22 minutes of yelling and threateningly staring at the neurons had a deleterious effect upon their self esteem.  This experiment would aim to correct for that effect. 
\end{itemize}
\end{multicols}

\section{\\SUMMARY}
\begin{multicols}{2}
We have offered a magic-based alternative to the biophysical computational modeling techniques that are common in neuroscience, which we hope will serve as a refreshing point of comparison as we move forward.  Trends wax and wane, and biophysical realism might not be all that it is currently cracked up to be. 

The work we have presented thus far brings to mind many questions, which we shall address in a series of follow-up papers.  For example, how would results compare upon repeated and uncorrelated samplings of random connectivities?  How would they compare across different brands of pipe cleaner?  And what if we were to switch up all the colors assigned to the neurons' constituent compartments?  

Finally, to conclude on an auspicious note, we make an extremely brief and cryptic statement regarding our intended future work.      
\end{multicols}  

\section{\\ACKNOWLEDGEMENTS}
\begin{multicols}{2}
The materials used in this research were generously donated by Ms. Clarissa Rosenberg, this author's eight-year-old next-door neighbor.   
\end{multicols}

\bibliographystyle{acm}
\nocite{*}
\bibliography{bibliography}
\end{document}